# Three-Dimensional Electrostatic and Quantum-Confinement Modeling of Silicon Nanowire Double Quantum Dots

# (Version – 2)

Nilesh Pandey[1], Dipanjan Basu[2], Leonard F Register[1] and Sanjay K Banerjee[1]

[1]University of Texas at Austin, Austin, USA, [2]Synopsys Inc., Hillsboro, OR, USA

email: pandey@utexas.edu

*Abstract* - We present a three-dimensional simulation study of silicon nanowire double quantum dots (DQDs) with leads at $T$ = 2 K, which extends beyond traditional effective mass or quasi-1D and quasi-2D approaches typically applied to bulk or planar geometries. A 3-D Poisson solver is self-consistently coupled to 2-D Schrodinger along slices normal to transport (width × thickness) to obtain spatially varying subbands and wavefunctions at $T$ = 2 K. The slice approximation is justified by the large aspect ratio ($L_{tot}/W$ > 20) and by the small (< 1.2%) wavefunction variation observed along the transport direction. The resulting effective conduction-band profile is imported into a full-wave, open-boundary Schrodinger solver to compute the transmission spectrum [$T$(E)], and the tunnel coupling ($t_c$) is evaluated from the bonding–antibonding splitting of the first two resonances in $T$(E). The simulations show that narrow dots ($W \sim 5$ nm) provide strong confinement and robust single-electron localization but require higher plunger-gate voltage, whereas wider dots ($W \sim 20$ nm) load electrons at lower bias but form shallower, more delocalized states. The tunnel coupling decreases as the dot width and length are increased, due to the reduced wavefunction overlap between the dots, and saturates once W ≥ $2L_{PG}$, when longitudinal confinement is dominated by the plunger gate length. The simulated tunnel coupling trend agrees with experimental data reported for the Si DQD device.

## I. Introduction

Silicon-based quantum bits (qubits) have emerged as a leading platform for scalable quantum computing owing to their compatibility with CMOS fabrication, long coherence times, and mature materials technology [1]–[5]. Qubits implemented in silicon can be broadly classified into two types: spin-based and charge-based. In spin qubits, the information is encoded in the electron spin state, offering excellent coherence but slower manipulation speeds [1]- [5]. In contrast, charge qubits encode quantum information in the spatial localization of single electrons within a double quantum dot (DQD) potential, where the logical states |0⟩ and |1⟩ correspond to the electron occupying the left or right dot, respectively [6]-[7]. The electron wavefunction overlap between the two dots determines the tunnel coupling, which controls the exchange interaction, qubit operation frequency, and charge hybridization dynamics [7]-[9].

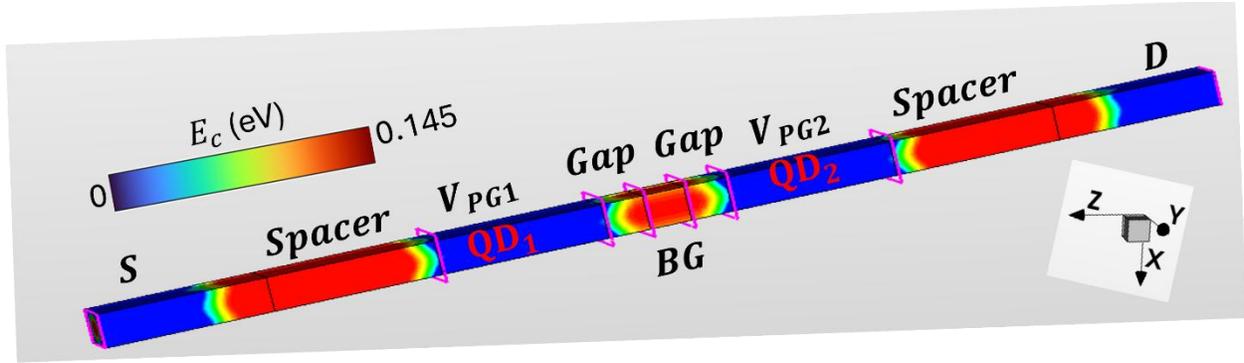

**Fig. 1**. 3-D nanosheet Si Qubit butid in TCAD SPROCESS. Parameters: Gap oxide length between plunger & barrier gate = 5 nm, Gate oxide thickness ($SiO_2$) = 1 nm, Nanosheet thickness ($H$) = 5 nm, Barrier gate length ($L_{BG}$) = 5 nm, gate metal work function = 4.36 eV, Spacer length = 20 nm, Plunger gate length: mentioned in each figure, $V_{PG1}$ = $V_{PG2}$ = 0.4 V, $V_{BG}$ = 0.3 V, and $V_{ds}$ = 0 V.

In the DQD system, the charge states $|1, 0\rangle$ and $|0, 1\rangle$ are separated by a potential barrier that suppresses electron tunneling to the source and drain via the Coulomb blockade, while allowing interdot tunneling when the barrier between dots is tuned electrostatically [7]–[10]. When tunneling to the leads is blocked but coupling between the two dots is finite, the electron wavefunction hybridizes across both wells, forming coherent superposition states $|1, 0\rangle + |0, 1\rangle$ that define the qubit basis [7]. During readout or initialization, the barriers to the leads are momentarily lowered to allow controlled tunneling, which collapses the superposition and enables charge detection through nearby sensors [10]-[14]. The ability to maintain this coherent hybridization between two Coulomb-blockaded charge states provides the foundation for quantum logic operations and enables parallel quantum computation in the Hilbert space, vastly exceeding classical processing [7]-[14].

On the modeling side, multiphysics and TCAD-style solvers have been developed for Si qubit structures, including Schrödinger–Poisson treatments, valley-splitting and single-electron transistor (SET)- based analyses, and statistical/wavefunction studies [15]–[20]. Furthermore, quantum-simulation work spans Hubbard-model realizations and digital/analog simulation frameworks applicable to coupled-dot qubits [21]–[26].

Previous work from IMEC combined multiphysics models to address stress, micromagnetics, valley splitting, and spin dynamics, but their treatment of confinement was bulk-like and neglected wavefunction evolution along the width [15]. However, a real Qubit will always be in 3-D [2], [9]. In our earlier work [28], we employed a 2-D Poisson and 1-D Schrödinger framework to study electrostatics and estimate basic tunnel coupling. However, these approaches are not representative of the 3-D quantum confinement and only valid for 2-D geometry. In contrast, our work employs a full 3-D Poisson and 2-D Schrödinger TCAD approach that directly captures confinement in all cross-sectional directions, reveals width-dependent wavefunction splitting and higher-mode formation, and establishes clear geometry-driven trends in tunnel coupling.

The present work introduces a comprehensive quantum transport framework that solves for subband transport, incorporating 2-D quantum confinement perpendicular to the transport direction and quasi-3-D confinement in quantum dots, coupled self-consistently with the full 3-D Poisson equation. This approach captures (i) Captures the confinement in all cross-sectional directions, (ii) exposes width-dependent wavefunction splitting and higher-mode formation, and (iii) links these effects to clear geometry-driven trends in tunnel coupling. This work models a subband-transport framework self-consistently coupled with 3-D electrostatics at $T = 2$ K (compared to $\sim 10$ K in [15]), focusing on how geometrical confinement influences interdot tunnel coupling in CMOS-compatible Si qubits. Note that band-structure refinements (e.g., valley splitting, valley-orbit coupling) are outside the present scope, as our focus is on the full qubit device (> 100 nm, including spacers, and S/D).

## II. 3-D SENTAURUS QTX MODELING

Fig. 1(a) shows the conduction-band profile of the simulated 3-D Si nanowire qubit device generated in TCAD SPROCESS. The Si nanowire is wrapped by a 1 nm SiO2 layer, above which plunger and barrier gates define the quantum dots and interdot barrier.

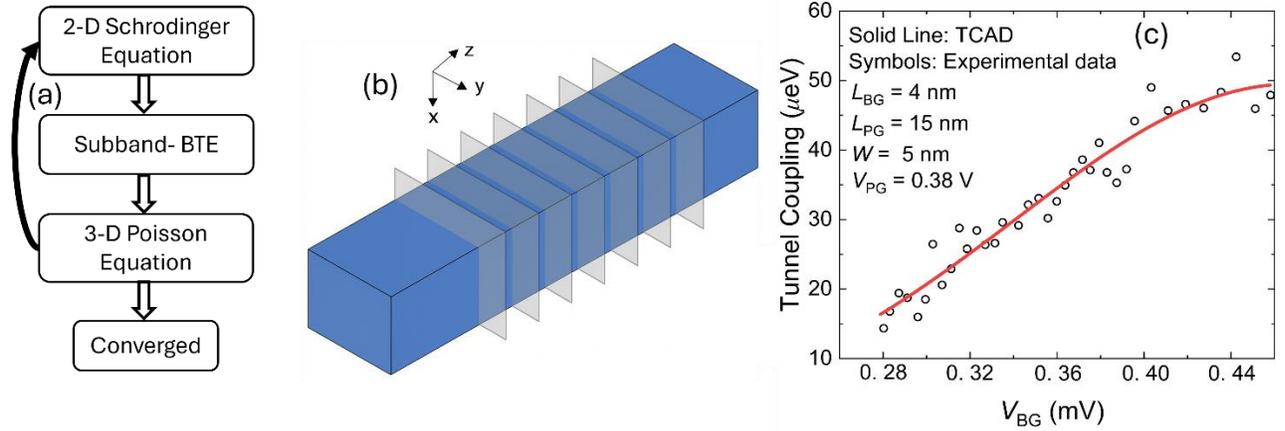

**Fig. 2** (a,b) Self-consistent quantum-transport framework implemented using Synopsys Sentaurus QTX, where a 3-D Poisson solver is coupled with 2-D Schrödinger along the width × thickness slices (≈ 0.1 nm spacing) to obtain spatially varying subband edges and wavefunctions. (c) Validation of the TCAD simulation results against the experimental data reported in [27]. The interdot tunnel coupling, ($t_c$), increases quasi-linearly with the barrier-gate voltage ($V_{BG}$) and saturates as the quantum dots begin to merge, consistent with the trends observed in [27]. The extracted coupling values, $t_c \sim 10 - 50$ $\mu eV$ confirm an overlap-controlled origin of tunneling between the dots.

Propagating quantum states are obtained self-consistently using Synopsys SENTAURUS QTX, where a 3-D Poisson solver is coupled with 2-D Schrödinger along the width × thickness slices (≈ 0.1 nm spacing) normal to the transport direction, as shown in **Fig. 2(a,b)**. The slice-wise bound levels define spatially varying subband edges $E_n(z)$ and wavefunctions $\psi(x, y, z)$), which construct the longitudinal conduction-band profile for quantum-transport calculations. This approach captures confinement and electrostatics self-consistently. Scattering is neglected, which is appropriate for near-equilibrium cryogenic operation. The effective conduction band energy $E_{net}(z) = E_c(z) + \Delta E_n(z)$ is imported into COMSOL to solve the open-boundary Schrodinger equation as a full-wave scattering problem (see section IV).

**Fig. 2(c)** validates the TCAD results against experimental data [27]. The interdot tunnel coupling ($t_c$) increases quasi-linearly with barrier-gate voltage $V_{BG}$ and saturates as the dots merge, consistent with measurements. The extracted tc ~ 10–50 µeV confirms wavefunction overlap between the dots. The $V_{BG}$ range (here 0.28–0.44 V vs. ~ 3.4–3.85 V in [27]) arises from geometry and work-function differences, but the functional trend and energy scale remain consistent. Because experimentally reported $t_c$ spans orders of magnitude [9], [27], [29], exact numerical matching is less critical, and reproducing the measured trend in [27] provides sufficient physical validation of the TCAD framework.

### III. Justification of the 2-D Schrödinger Approximation

In the present framework, Schrödinger's equation is solved in the width–thickness (x–y) cross-section at each slice along the transport direction, while the electrostatics are fully resolved in three dimensions through the Poisson equation. This approximation is justified by the large aspect ratio between the transport length and the confined dimensions. For example, even at the widest channel width considered (W = 25 nm, H = 5 nm), the source–drain transport direction exceeds the confined dimension by more than a factor of six, and for narrower dots the ratio is greater than twenty. As a result, the transverse wavefunction is expected to vary only slightly along the transport direction.

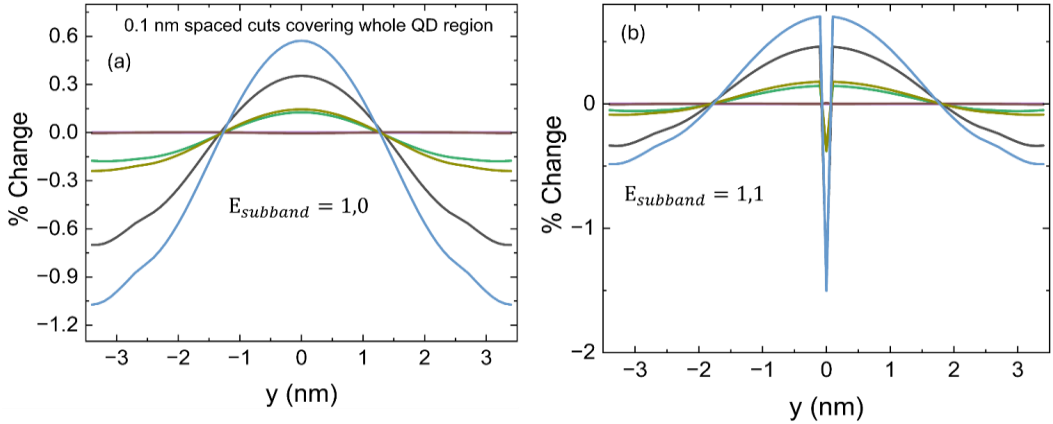

**Fig. 3.** Percentage change in the wavefunction along the channel (z-) direction for (a) the first subband ($E_{subband}$ = 1,0) and (b) the second subband ($E_{subband}$ = 1,1). The z-cuts are spaced at 0.1 nm, corresponding to parallel width × thickness slices through the quantum dot. For each slice, a one-dimensional profile along the width (y-) direction is extracted. The variation is negligible (<1.2% for the 1,0 mode and <2% for the 1,1 mode), confirming that the transverse wavefunction remains nearly constant along the transport direction and validating the 2-D Schrödinger approximation.

**Fig. 3.** Percentage change in the wavefunction along the channel (z) direction with respect to the center of the quantum dot. The z-cuts are spaced at 0.1 nm, corresponding to successive slices of width × thickness parallel to the channel direction. For each slice, a 1-D profile along the width ($y$) direction is extracted, and the relative percentage change is plotted. The variation is negligible (< 1.2 % for the 1,0 mode and < 2 % for the 1,1 mode), indicating that the wavefunction remains nearly constant along the $z$-direction within the dot region. This validates the assumption that the Schrödinger equation can be solved only in the width–thickness plane while treating transport separately. These results confirm that treating transport separately while resolving quantum confinement only in the transverse plane is physically well justified.

## IV. Quantum Transport (Tunnel Coupling) Modeling

From the SBTE (3-D Poisson + 2-D Schrödinger) solution, the transverse ground subband wavefunction shows negligible change between slices separated by 0.1 nm within the quantum dot (see **Fig. 3**), with only gradual modulation across the barrier. Because the transport direction between source and drain is much longer than the transverse dimensions ( ~ > 20 times), treating the transverse wavefunction as nearly constant along transport is justified. The effective longitudinal potential $V(z)$, constructed directly from the QTX-derived conduction-band profile $E_c(z)$ and subband offset $\Delta E_n(z)$, is then used in COMSOL to solve the stationary Schrödinger equation with open boundaries as a full-wave scattering problem by following the equations [28], [30]. The effective conduction band energy is given by

$$E_{net}(z) = E_c(z) + \Delta E_n(z) \quad (1)$$

which is imported into COMSOL Multiphysics to solve the open-boundary Schrödinger equation as a full-wave scattering problem.

The effective wavevector is defined as

$$|k| = \frac{\sqrt{2m_e^*\{E - E_{net}(z)\}}}{\hbar}$$

and applied at the domain boundaries through $n.\nabla.\psi = -ik\psi$. This full-wave formulation allows COMSOL to compute the transmission spectrum $T(E)$, and the tunnel coupling ($t_c$) is formally defined from the energy splitting of the first two resonant transmission peaks, $E_0$ and $E_1$, as

$$t_c = \frac{E_0 - E_1}{2} = \frac{\Delta E}{2}$$

where $E_0$ and $E_1$ correspond to the bonding and antibonding states of the double quantum dot.

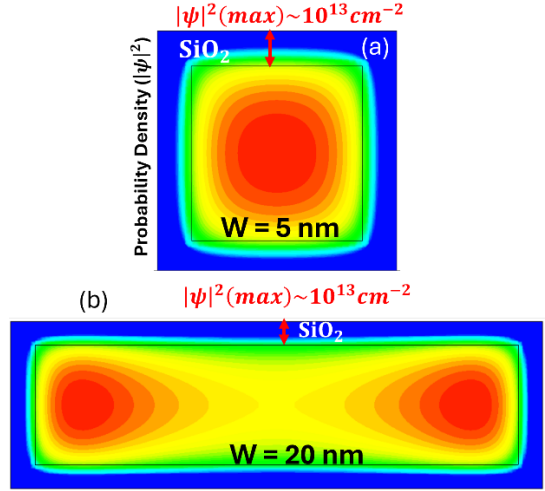

**Fig. 4.** (a) Electron probability density ($|\psi|^2$) in the $W \times H$ cross-section for z cut at the center of the dot ($Z = 0$), highlighting strong quantum confinement and wavefunction localization at the channel center for $W =5$ nm. (b) With increasing channel width ($W=20$ nm ), quantum confinement in the Si nanosheet is significantly reduced compared to the ultra-narrow ($W=5$ nm) case. (a) The ground-state wavefunction is no longer centered but instead forms two pronounced maxima near the sidewalls, showing that electrons preferentially localize close to the Si/SiO$_2$ interfaces. (b) This weaker confinement results in carrier density peaks near the Si/SiO$_2$ surfaces. See Fig. 1 for the remaining parameters.

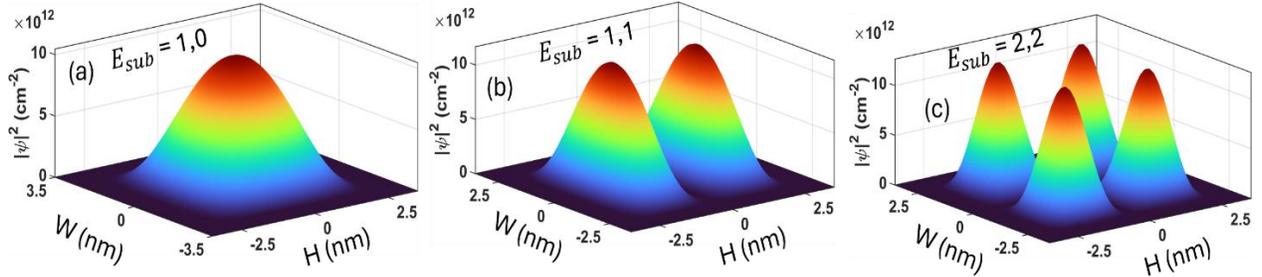

**Fig.5** Numerically calculated probability densities ($|\psi|^2$ for the lowest three subbands [1,0], [1,1], and [1,3] in the plunger gate cross-section. Each plot shows the quantum dot's width and thickness directions, with quantum numbers denoting the number of nodes in the thickness and width, respectively. Progression to higher subbands reveals more oscillatory nodes, reflecting stronger quantum confinement and higher subband energies. This quantization governs subband splitting and wavefunction overlap, which are crucial for tunnel coupling and qubit performance. **Rest parameters:** Fig. 1 (applied biases).

## V. WAVEFUNCTION DYNAMICS

**Fig. 4** shows spatial profiles of the ground-state probability density $|\psi^2|$ vs. energy in the quantum dot for two dot widths. For $W = 5$ nm (a), strong quantum confinement leads to a single, centrally peaked wavefunction, reflecting tight localization and large subband spacing. Although the electric-field-induced potential well is deepest at the interface, quantum confinement dominates, resulting in electron density being peaked at the center, rather than at the interface as would be expected classically. In contrast, for $W$=20 (b), the wavefunction spreads toward the sides of the channel, forming two distinct maxima near the sidewalls. This reduced confinement lowers the ground-state energy and subband separation, allowing higher electron occupancy at lower gate voltages at the itnerface. **Fig. 5** shows the calculated probability densities for the lowest three transverse subbands, labeled as [1,0], [1,1], and [2,2], in the plunger gate cross-section. Each probability density is plotted across the width and thickness directions, with the quantum numbers ($n_x$,$n_y$) corresponding to the number of nodes along the thickness ($n_x$) and width ($n_y$) of the quantum dot, respectively. The progression from [1,0] to [2,2] illustrates the emergence of oscillatory nodes in the transverse profile.

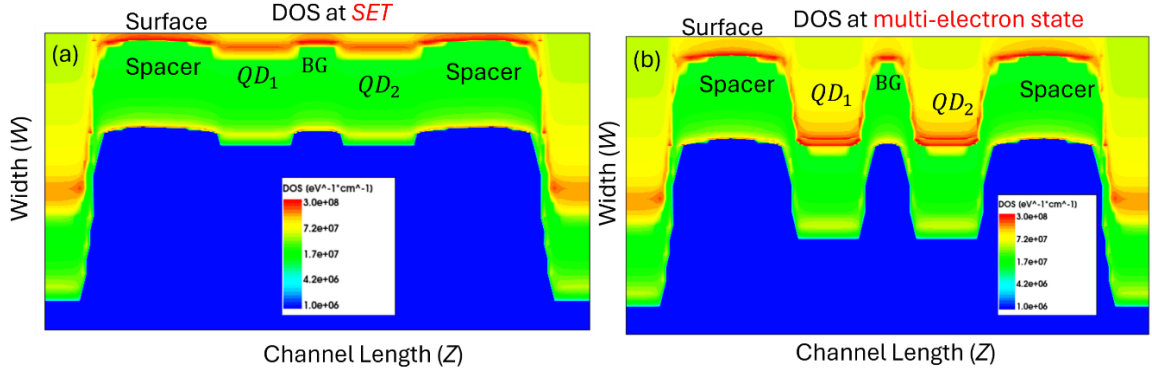

**Fig. 6** Density of state (DOS) maps showing quantum dot (QD) confinement at (a) single-electron dot and (b) multi-electron regimes. In a single-electron dot, the QD depth is ~19 meV (≫ $k_B$T), supporting sharp subband DOS peaks and strong electron localization. Multi-electron regime features deeper wells (~150 meV) and multiple subbands in the QD. The DOS and subbands (negligible) in spacers are much lower, ensuring confinement of the electron within the QD. Furthermore, the shallower well (single electron state) is ~ 100 times deeper than the thermal energy at $T$ = 2 K, again ensuring that the electron state in the well remains trapped. The continuous atmospheric level DOS also indicates that the model does not employ effective mass approximations and is therefore accurate for quantum simulations. **Rest parameters:** Fig. 1 (applied voltages, etc.).

## VI. SINGLE ELECTRON STATE V/S MULTI-ELECTRON DOT

**Fig. 6** shows the spatial distribution of the density of states (DOS) in the *channel × width* ($Z \times Y$) direction of a double quantum dot device. The DOS illustrates the transition from single-electron to multi-electron regimes. In the single-electron case (a), the quantum dot (QD) well depth is ~19 meV but still over 100 times the thermal energy at $T$ = 2 K, so even a shallow well robustly traps the electron, with occupation primarily in the lowest subband. For higher plunger voltages (b), the QD depth increases (~150 meV), and multiple occupied subbands appear, signaling multi-electron occupation. Although there is some probability of finding carriers in the spacer regions due to tunneling, electron localization remains strong in the QD due to a still quite high well binding energy relative to $k_B T$, > 100 times.

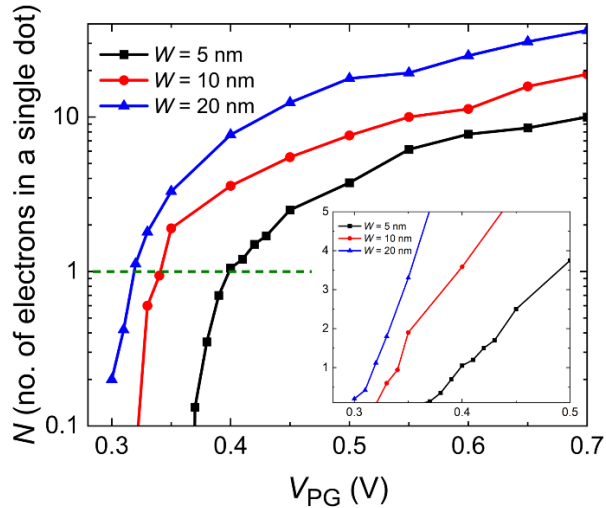

**Fig. 7.** Number of electrons ($N$) in a single quantum dot as a function of plunger gate voltage ($V_{PG}$) for the various channel widths. For narrower channels ($W$ = 5 nm), stronger quantum confinement and larger subband spacing require higher $V_{PG}$ to accumulate the first electron. In contrast, for wider channels ($W$ = 20 nm), the onset of single-electron occupancy and further electron addition occur at lower $V_{PG}$, reflecting reduced quantization and Coulomb energies. The leftward shift in the single-electron dot transition with increasing width directly correlates with the spatial delocalization of the wavefunction, as shown in Fig. 3, underscoring how device geometry dictates electrostatic tunability and quantum dot operation. **Rest parameters:** Fig. 1 (applied voltages, etc.).

**Fig. 7** shows the number of electrons in the regions beneath one plunger gate as a function of plunger gate voltage ($V_{PG}$) for three dot widths ($W$). The narrower the dot, the higher the $V_{PG}$ required to accumulate a single electron per plunger gate and the subsequent electrons. In contrast, for wider channels ($W$ = 20 nm), the transition to the single electron regime and subsequent electron additions occur at lower values of $V_{PG}$, respectively. This shift in the $V_{PG}$ with channel width, however, is tied to both quantum confinement (**Figs. 4 & 5**) and increasing electrostatics.

Electrostatic repulsion for a given total charge increases with reduced channel width following $E_{QC} \propto 1/W^2$. For the latter, greater confinement also increases electrostatic repulsion for even a single electron, which is an artifact of not yet including a counterbalancing exchange potential, adding somewhat to the required voltage to reach single electron occupation, but does not compromise the essential physics discussed. For the two wider, less quantum-confined dots, lower gate voltages are required to reach one electron per gate. However, as the width increases and the barrier voltage is kept fixed, the well becomes shallower and electrons delocalize. Wider dots reach one-electron-per-dot at lower gate voltages, but without forming a double quantum dot as shown in **Fig. 8**.

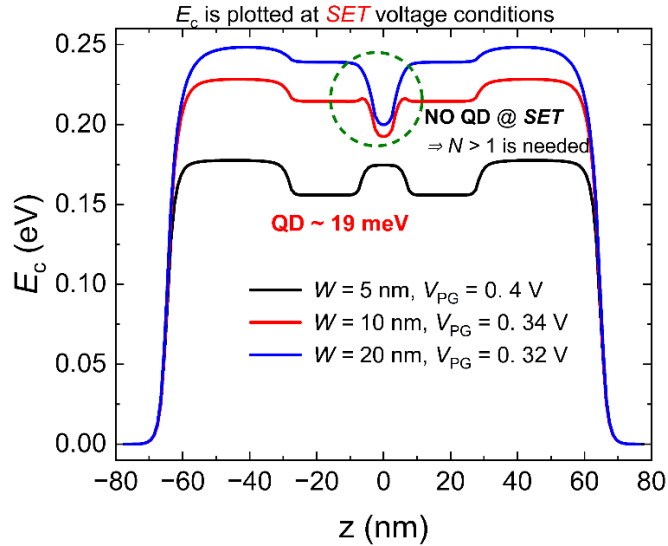

**Fig. 8.** Conduction band profiles ($E_c$) along the channel ($Z$) direction for three dot widths at the single-electron dot condition with their respective threshold plunger gate voltage ($V_{PG}$). As the width increases, the required $V_{PG}$ for electron occupancy decreases, but the quantum well becomes progressively shallower and eventually vanishes, leading to electron delocalization [ (a), (b) & (c)]. For larger widths, the absence of a well-defined quantum dot makes robust single-electron localization challenging, highlighting the inherent trade-off between electrostatic tunability and quantum confinement in device design. **Rest parameters:** Fig. 1 (applied voltages, etc.).

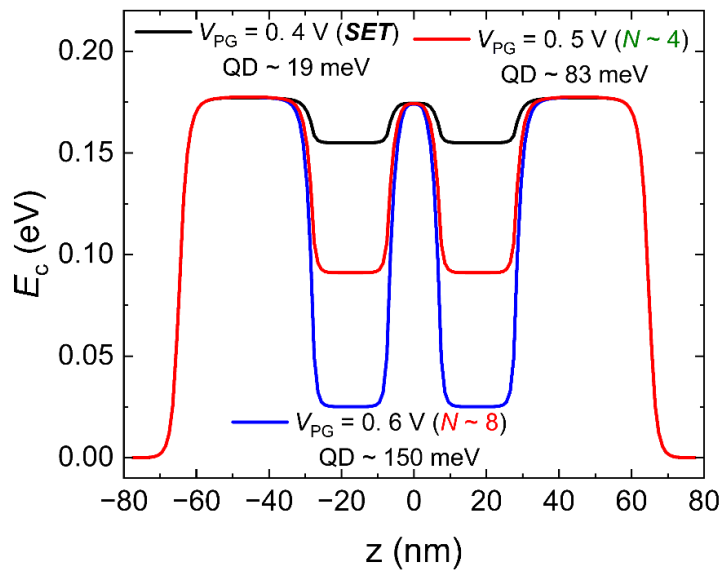

**Fig. 9** The evolution from single- to multi-electron quantum dot regimes. As $V_{PG}$ increases, the quantum dot well deepens substantially, enhancing electron confinement and robustness to electrical noise. However, this deeper well also enables multi-electron occupation, introducing additional Coulomb interactions and charge noise that degrade the single-electron state and qubit performance. This highlights the trade-off between noise immunity and maintaining a true single-electron quantum dot. **Rest parameters:** Fig. 1 (applied voltages, etc.)

**Fig. 9** shows that a deeper quantum well improves robustness against thermal and electrical noise but also allows higher-energy electron states (multi-electron) to be occupied, leading to additional Coulomb interactions and coupling-induced noise. This degrades coherence and reduces the purity of the single-electron regime required for optimal qubit performance.

## V. III *TUNNEL COUPLING*

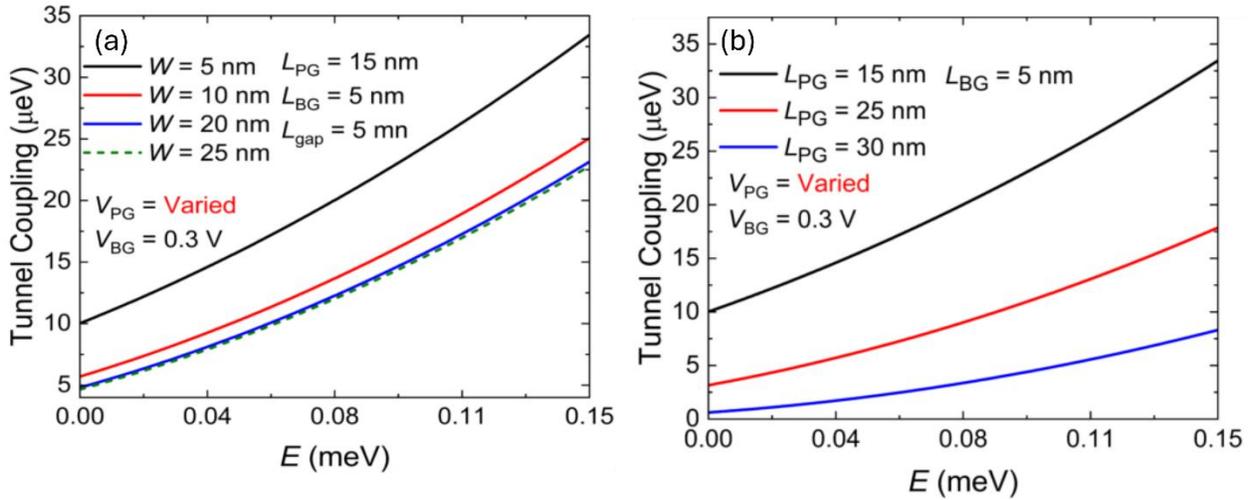

**Fig. 10** (a) Reducing channel width (W) raises the transverse subband energy ($\propto 1/W^2$), effectively lowering the interdot barrier and enhancing the tunnel coupling, which saturates once W ≥ 2$L_{PG}$, where longitudinal scaling ($L_{PG}$) dominates. (b) Increasing plunger-gate length ($L_{PG}$) enlarges dot separation and wavefunction decay, reducing tunnel coupling.

**Fig. 10(a)** plots the tunnel coupling versus energy. The plunger-gate voltage ($V_{PG}$) is adjusted for each width to maintain approximately constant dot depth (≈ 20 meV). Energies are referenced to the dot-well minimum. As W decreases, stronger transverse confinement lowers the effective interdot barrier ($E_\perp \propto 1/W^2$) and enhances left/right wavefunction overlap, yielding larger tunnel coupling. Within any given width, tunnel coupling rises with the energy (E) as the state approaches the barrier top, leading to an effective reduction in the net barrier for wavefunction overlap. For sufficiently larger $W$ ($widht \geq 2 \times Plunger\ gate\ length$), the curves converge because the longitudinal barrier is then governed mainly by plunger gate geometry $(\propto 1/L_{PG}^2)$
**Fig. 10(b)** shows that increasing plunger gate length ($L_{PG}$) enlarges the interdot separation and barrier width, thereby strengthening evanescent decay across the barrier and reducing wavefunction overlap; consequently, tunnel coupling decreases.

## IX. CONCLUSION

We present a 3-D electrostatics and 2-D Schrödinger-slice framework for silicon nanowire double quantum dots, using the resulting longitudinal potential in a full-wave scattering model to compute tunneling probability and extract tunnel coupling from the first two resonant peaks at $T = 2\,K$. The simulations reveal that narrower quantum dots yield deeper quantum wells and stronger localization, leading to robust single-electron operation but requiring higher plunger bias for loading. Wider dots lower the gate threshold but weaken confinement, producing shallower wells and multi-electron occupation. The tunnel coupling ($t_c$) increases as $W$ and dot length ($L_{PG}$) decrease due to enhanced interdot overlap and saturates once $W \geq 2L_{PG}$. These results establish geometry-driven design rules for optimizing well depth, tunnel coupling, and single-electron control in CMOS-compatible Si qubits.

**FUTURE SCOPE**

The workflow is designed to be single-particle and cryogenic/near-equilibrium; phonons, disorder, and explicit spin dynamics are left for future work. Even so, the established geometry trends offer an immediate guide to co-optimizing confinement, valley splitting, and interdot coupling in scaled devices.

Acknowledgment: This work was supported by NSF National Nanotechnology Coordinated Infrastructure (NNCI) (Grant Number: ECCS2025227)

# References

[1] D. Loss and D. P. DiVincenzo, "Quantum computation with quantum dots," Physical Review A, vol. 57, pp. 120–126, 1998, doi: 10.1103/Phys-RevA.57.120.

[2] R. Hanson, L. P. Kouwenhoven, J. R. Petta, S. Tarucha, and L. M. K. Vandersypen, "Spins in few-electron quantum dots," Reviews of Modern Physics, vol. 79, pp. 1217–1265, 2007, doi: 10.1103/RevMod-Phys.79.1217.

[3] R. Maurand, X. Jehl, D. Kotekar-Patil, A. Corna, H. Bohuslavskyi, R. Laviéville, L. Hutin, S. Barraud, M. Vinet, M. Sanquer, and S. De Franceschi, "A cmos silicon spin qubit," Nature Communications, vol. 7, no. 1, p. 13575, Nov 2016, doi: 10.1038/ncomms13575.

[4] P. Steinacker, N. Dumoulin Stuyck, W. H. Lim, T. Tanttu, M. Feng, S. Serrano, A. Nickl, M. Candido, J. D. Cifuentes, E. Vahapoglu, S. K. Bartee, F. E. Hudson, K. W. Chan, S. Kubicek, J. Jussot, Y. Canvel, S. Beyne, Y. Shimura, R. Loo, C. Godfrin, B. Raes, S. Baudot, D. Wan, A. Laucht, C. H. Yang, A. Saraiva, C. C. Escott, K. De Greve, and A. S. Dzurak, "Industry-compatible silicon spin-qubit unit cells exceeding 99% fidelity," Nature, vol. 646, no. 8083, pp. 81–87, Oct 2025, doi: 10.1038/s41586-025-09531-9.

[5] B. E. Kane, "A silicon-based nuclear spin quantum computer," Nature, vol. 393, no. 6681, pp. 133–137, May 1998, doi: 10.1038/30156.

[6] W. G. Van der Wiel, S. De Franceschi, J. M. Elzerman, T. Fujisawa, S. Tarucha, and L. P. Kouwenhoven, "Electron transport through double quantum dots," Reviews of Modern Physics, vol. 75, pp. 1–22, 2002, doi: 10.1103/RevModPhys.75.1.

[7] T. Hayashi, T. Fujisawa, H. D. Cheong, Y. H. Jeong, and Y. Hirayama, "Coherent manipulation of electronic states in a double quantum dot," Physical Review Letters, vol. 91, p. 226804, 2003, doi: 10.1103/Phys-RevLett.91.226804.

[8] G. Burkard, D. Loss, and D. P. DiVincenzo, "Coupled quantum dots as quantum gates," Physical Review B, vol. 59, pp. 2070–2078, 1999, doi: 10.1103/PhysRevB.59.2070.

[9] F. A. Zwanenburg, A. S. Dzurak, A. Morello, M. Y. Simmons, L. C. L. Hollenberg, G. Klimeck, S. Rogge, S. N. Coppersmith, and M. A. Eriksson, "Silicon quantum electronics," Rev. Mod. Phys., vol. 85, pp. 961–1019, Jul 2013, doi: 10.1103/RevModPhys.85.961.

[10] L. K. Grover, "Quantum mechanics helps in searching for a needle in a haystack," Phys. Rev. Lett., vol. 79, pp. 325–328, Jul 1997, doi:10.1103/PhysRevLett.79.325.

[11] F. Arute, K. Arya, R. Babbush et al., "Quantum supremacy using a programmable superconducting processor," Nature, vol. 574, no. 7779, pp. 505–510, 2019, doi: 10.1038/s41586-019-1666-5.

[12] D. M. Zajac, A. J. Sigillito, M. Russ, F. Borjans, J. M. Taylor, G. Burkard, and J. R. Petta, "Resonantly driven cnot gate for electron spins," Science, vol. 359, pp. 439–442, 2018, doi: 10.1126/science.aao5965.

[13] L. DiCarlo, H. J. Lynch, A. C. Johnson, L. I. Childress, K. Crockett, C. M. Marcus, M. P. Hanson, and A. C. Gossard, "Differential charge sensing and charge delocalization in a tunable double quantum dot," Phys. Rev. Lett., vol. 92, p. 226801, Jun 2004, doi: 10.1103/Phys-RevLett.92.226801.

[14] K. D. Petersson, J. R. Petta, H. Lu, and A. C. Gossard, "Quantum coherence in a one-electron semiconductor charge qubit," Phys. Rev. Lett., vol. 105, p. 246804, Dec 2010, doi: 10.1103/PhysRevLett.105.246804.

[15] F. A. Mohiyaddin, G. Simion, N. I. D. Stuyck, R. Li, F. Ciubotaru, G. Eneman, F. M. Bufler, S. Kubicek, J. Jussot, B. Chan, T. Ivanov, A. Spessot, P. Matagne, J. Lee, B. Govoreanu, and I. P. Raduimec, "Multiphysics simulation

design of silicon quantum dot qubit devices," in 2019 IEEE International Electron Devices Meeting (IEDM), 2019, pp. 39.5.1–39.5.4, doi: 10.1109/IEDM19573.2019.8993541.

[16] M. M. E. K. Shehata, G. Simion, R. Li, F. A. Mohiyaddin, D. Wan, M. Mongillo, B. Govoreanu, I. Radu, K. De Greve, and P. Van Dorpe, "Modeling semiconductor spin qubits and their charge noise environment for quantum gate fidelity estimation," Phys. Rev. B, vol. 108, p. 045305, Jul 2023, doi: 10.1103/PhysRevB.108.045305.

[17] L. Bourdet and Y.-M. Niquet, "All-electrical manipulation of silicon spin qubits with tunable spin-valley mixing," Phys. Rev. B, vol. 97, p. 155433, Apr 2018, doi: 10.1103/PhysRevB.97.155433.

[18] J. K. Gamble, *et al.*, "Valley splitting of single-electron si mos quantum dots," Applied Physics Letters, vol. 109, no. 25, p. 253101, 12 2016, doi: 10.1063/1.4972514.

[19] B. Venitucci and Y.-M. Niquet, "Simple model for electrical hole spin manipulation in semiconductor quantum dots: Impact of dot material and orientation," Phys. Rev. B, vol. 99, p. 115317, Mar 2019, doi: 10.1103/PhysRevB.99.115317.

[20] M. F. Gonzalez-Zalba, S. de Franceschi, E. Charbon, T. Meunier, M. Vinet, and A. S. Dzurak, "Scaling silicon-based quantum computing using cmos technology," Nature Electronics, vol. 4, no. 12, pp. 872–884, Dec 2021, doi: 10.1038/s41928-021-00681-y.

[21] D. Jaksch, C. Bruder, J. I. Cirac, C. W. Gardiner, and P. Zoller, "Cold bosonic atoms in optical lattices," Phys. Rev. Lett., vol. 81, pp. 3108–3111, Oct 1998, doi: 10.1103/PhysRevLett.81.3108.

[22] T. Hensgens, T. Fujita, L. Janssen, X. Li, C. J. Van Diepen, C. Reichl, W. Wegscheider, S. Das Sarma, and L. M. K. Vandersypen, "Quantum simulation of a fermi–hubbard model using a semiconductor quantum dot array," Nature, vol. 548, no. 7665, pp. 70–73, Aug 2017, doi:10.1038/nature23022.

[23] J. Salfi, J. A. Mol, R. Rahman, G. Klimeck, M. Y. Simmons, L. C. L. Hollenberg, and S. Rogge, "Quantum simulation of the hubbard model with dopant atoms in silicon," Nature Communications, vol. 7, no. 1, p. 11342, Apr 2016, doi: 10.1038/ncomms11342.

[24] M. A. Nielsen and I. L. Chuang, Quantum Computation and Quantum Information: 10th Anniversary Edition. Cambridge University Press, 2010, doi: https://doi.org/10.1017/CBO9780511976667.

[25] G. Ortiz, J. E. Gubernatis, E. Knill, and R. Laflamme, "Quantum algorithms for fermionic simulations," Phys. Rev. A, vol. 64, p. 022319, Jul 2001, doi: 10.1103/PhysRevA.64.022319.

[26] I. M. Georgescu, S. Ashhab, and F. Nori, "Quantum simulation," Rev. Mod. Phys., vol. 86, pp. 153–185, Mar 2014, doi: 10.1103/RevMod- Phys.86.153.

[27] H. G. J. Eenink, L. Petit, W. I. L. Lawrie, J. S. Clarke, L. M. K. Vandersypen, and M. Veldhorst, "Tunable coupling and isolation of single electrons in silicon metal-oxide-semiconductor quantum dots," Nano Letters, vol. 19, no. 12, pp. 8653–8657, 2019, doi: 10.1021/acs.nanolett.9b03254.

[28] N. Pandey, D. Basu, Y. S. Chauhan, L. F. Register, and S. K. Banerjee, "Engineering si-qubit mosfet: Quantum-electrostatic integration at cryogenictemperatures," IEEE Transactions on Electron Devices, vol. 72, no. 7, pp. 3881–3888, 2025, doi: 10.1109/TED.2025.3570285.

[29] F. Borjans *et al.*, "Probing the variation of the intervalley tunnel coupling in a silicon triple quantum dot," PRX Quantum, vol. 2, p. 020309, Apr 2021, doi: 10.1103/PRXQuantum.2.020309.

[30] COMSOL Multiphysics Version: 6.1, COMSOL AB, Semiconductor Module Users Guide Stockholm, Sweden, 2022